\documentclass[useAMS,usenatbib,usegraphicx]{mn2e}
\usepackage{amssymb}
\newcommand\nH{n_\mathrm{H}}
\newcommand\percc{\mathrm{cm}^{-3}}
\newcommand\persqcm{\mathrm{cm}^{-2}}

\title[Search for OH 6 GHz maser emission]{A search for OH 6 GHz maser
  emission towards supernova remnants}
\author[McDonnell, Wardle \& Vaughan]{Korinne E. McDonnell\thanks{E-mail:
korinnem@physics.mq.edu.au}, Mark Wardle and Alan E. Vaughan\\
Department of Physics, Macquarie University, Sydney, 2019, Australia}
\begin{document}


\pagerange{\pageref{firstpage}--\pageref{lastpage}} \pubyear{2008}

\maketitle

\label{firstpage}

\begin{abstract}
OH masers at 1720~MHz have proven to be excellent indicators of
interactions between supernova remnants and molecular clouds.  OH excitation
calculations suggest that the 6049~MHz OH maser line is excited for
higher column densities than for the 1720~MHz line.  Previous
observations and modelling of
1612, 1665 and 1667~MHz OH absorption and 1720~MHz OH masers indicated that the column densities in some
supernova remnants, $\sim10^{17}$~$\persqcm$, may be high enough for 6049~MHz OH masers to exist.  It is therefore a
potentially valuable indicator of remnant--cloud interaction.

We present excitation calculations predicting the formation of
6049~MHz OH masers and results of a survey using the Parkes Methanol Multibeam
receiver for 6049, 6035 and 6030~MHz OH masers towards 35 supernova remnants, a
star-forming region and 4 fields
in the Large and Small Magellanic Clouds.     

Two new sites of 6035 and 6030~MHz OH maser
emission associated with star-forming regions have been discovered, but no 6049~MHz masers
were detected to a brightness temperature limit of $\sim$~0.3~--~0.6~K,
even though modelling of the OH excitation suggests that maser
emission should have been detected.  Our upper-limits
indicate that the OH column density for a typical remnant $\lesssim10^{16.4}$~$\persqcm$,
which conflicts with observed and modelled column densities.  One possible explanation is that 6049~MHz OH masers may be more sensitive to velocity coherence than 1720~MHz OH masers under some conditions.    
\end{abstract}


\section{Introduction}
Interactions between supernova remnants and molecular clouds are
likely, as the short lifetime of massive stars mean that they do not move far from the molecular
cloud in which they form.  Once the star goes supernova the expanding remnant shocks the
molecular cloud, heating and compressing
it and changing the chemical composition of the gas.  These shock waves
contribute to the disruption of the cloud and may also trigger star
formation.  

Signatures such as kinematics and disturbed morphology
have been used as
the primary evidence of such interactions
\citep[e.g.][]{1981ApJ...245..105W,1991A&A...247..529R}, but the
interpretation was inconclusive due to confusion of the
projection along the line of sight.  The mixed-morphology class of
supernova remnants is also believed to be an indicator of interaction
with a molecular cloud.  Mixed-morphology remnants appear
center-filled in X-rays and shell-like at radio wavelengths \citep{1998ApJ...503L.167R}.  The
supernova shock propagating into dense and possibly clumpy
interstellar gas is believed to be responsible for the interstellar
gas in the interior of the remnant, either through the evaporation of
clumps overrun by the supernova shock \citep{1991ApJ...373..543W} or heat
conduction reducing the internal temperature and density gradients
\citep{1992ApJ...401..206C,1999ApJ...511..798C,1999ApJ...524..179C,1999ApJ...524..192S}.     

The discovery of 1720~MHz OH masers in supernova remnants, created in the shocked region,
hence provided a solid diagnostic of the
interaction.  Produced by a transition within the ground rotational
state of OH, 1720~MHz OH line emission towards supernova remnants was first detected in W28 and
W44 by \citet{1968ApL.....2...81G}.  \citet*{1994ApJ...424L.111F} presented
high-resolution images of W28 that showed a more extensive
distribution of 1720~MHz emission than was expected and were the
first to suggest that the 1720~MHz OH maser line may be a powerful diagnostic of the
interaction of supernova remnants with molecular clouds.  \citet{1976ApJ...203..124E} showed
that collisions of OH molecules with H$_2$ molecules with kinetic
temperatures $T$ below 200~K strongly inverted the 1720~MHz line.  
Densities between 10$^3$ and 10$^5$ cm$^{-3}$ and temperatures between
50 and 125 K are needed for the inversion to function efficiently.  The OH column density needed is
10$^{16}$--10$^{17}$~$\persqcm$ \citep*[e.g.][]{1999ApJ...511..235L}.  At lower
column densities the opacity is too small and at higher column densities radiative
trapping thermalises the transition.  

Surveys of $\sim$200 galactic supernova remnants (SNRs), have found that 1 in 10 have 1720~MHz emission
\citep[e.g.][]{1997AJ....114.2058G,1998AJ....116.1323K}.  The masers
allow the kinetic distance to the SNR to be estimated as their velocities are close to the line-of-sight velocity of the
unshocked cloud complex and are consistent with other measures of the
systemic velocity of the remnant.  Zeeman observations allow the line-of-sight magnetic field
of the maser to be measured, from which the internal pressure of the
supernova remnant can be calculated \citep[e.g.][]{1997ApJ...489..143C}.  However, not all supernova remnants interacting with molecular clouds
will have 1720~MHz masers, as the conditions may not be correct to
produce 1720~MHz masers or the masers could be beamed away from the
line of sight \citep{1997AJ....114.2058G}.  So another
definitive signature of the interaction would be useful.

Masers resulting from a second maser transition would allow
constraints to be placed on physical conditions, such as density and
temperature in the masing region, particularly if the maser was
spatially coincident with an observed 1720~MHz maser.  It
would also confirm the Zeeman interpretation of the splitting seen
between the left- and right-hand circular polarisations (LHCP, RHCP)
at 1720~MHz, as circular polarisation can be produced by non-Zeeman
mechanisms (Elitzur 1996, 1998)\nocite{1996ApJ...457..415E,1998ApJ...504..390E}.

The 6049~MHz line in the first excited rotational level of OH is an analogue of the 1720~MHz satellite line.   OH excitation calculations \citep{2000ApJ...534..770P,2007IAUS..242..336W,2008ApJ...676..371P} suggest that the 6049~MHz satellite
line may be present at higher OH column densities where the 1720~MHz
line is weak or absent, and therefore might be a complementary signature of the
interaction of supernova remnants and molecular clouds.  The 4765~MHz
OH line in the second excited rotational state peaks at
an even higher column density of $N_{\mathrm{OH}}\sim
10^{18}$~$\persqcm$.  This column density is higher than that expected in
the interaction of supernova remnants and molecular clouds.  

This motivated us to conduct searches for 6049~MHz OH maser emission
using Parkes, ATCA, and exploratory time on the EVLA (ATCA and EVLA to
be reported in another paper).  Here we report
on the search using
the Parkes Radio Telescope and the Methanol Multibeam receiver.  The preliminary results of this
survey were presented in \citet*{2007IAUS..242..232M}.  No
6049~MHz maser emission was detected, however several 6030
and 6035~MHz OH masers associated with star-forming regions were
found.  It is not surprising that some star-forming regions were
unintentionally included in
the observed fields as star-formation can be triggered by supernovae.  \citet*{2007ApJ...670L.117F} also observed pointings
towards 14 supernova remnants at 6049~MHz but did not detect any OH
emission.  

Here we report on the final results of the Parkes survey for 6049, 6035 and 6030~MHz maser
emission towards a number of remnants.  Section \ref{sec:motivation}
outlines the theoretical model, Section \ref{sec:observations}
details the observations and data reduction, Section \ref{sec:results} presents the
results, which are discussed in Section \ref{discussion}.

\section[]{Inversion of satellite lines of OH}
\label{sec:motivation}
Here we present excitation calculations (previously discussed in \citealt{2007IAUS..242..336W}) showing that when the OH
column density exceeds $10^{17}$~$\persqcm$ at similar densities and
temperatures for 1720~MHz OH maser emission,
the inversion of the 1720~MHz line switches off, while the 6049~MHz transition in the first excited rotational state of OH becomes
inverted.  The calculations use a molecular excitation code developed by Wardle and tested against results from \citet{1999ApJ...511..235L}.   Similar results were found by \citet{2000ApJ...534..770P} and \citet{2008ApJ...676..371P}.

Our OH excitation calculations include the 32 lowest
energy levels, with transition wavelengths and A values from
\citet{1977A&A....60...55D} and \citet{1982ApJ...258..899B}, and rates for
collisional de-excitation by H$_2$ kindly provided by Alison Offer
(private communication).  Following \citet{1989ApJ...344..525L} and
\citet{1999ApJ...511..235L}, a uniform slab model was adopted for
the masing medium.  Radiative transfer is approximated using escape
probabilities based on the mean optical depth, including the effects
of line overlap.  The slab is parametrised by density $\nH$,
temperature $T$, column $N_\mathrm{OH}$, velocity width, and a
radiation field contributed by the Cosmic Microwave Background (CMB) and warm dust.

\begin{figure}\centering
 \includegraphics[width=84mm]{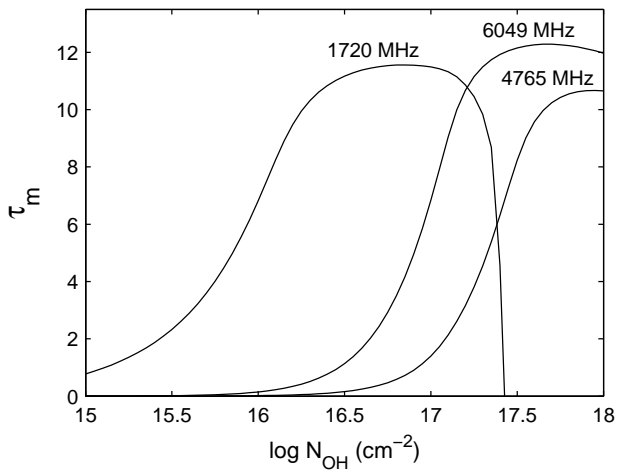}
  \caption{ Maser optical depth, $\tau_m$ ($I_\nu\propto
\exp(\tau_m)$)  of the 1720 MHz OH line
and its analogues at 6049 and 4765\,MHz, as a function of OH column density for T = 50\,K and
$\nH = 10^5 \percc$. }\label{fig:tau}
\end{figure}

\begin{figure}
    \centering
  \includegraphics[width=84mm]{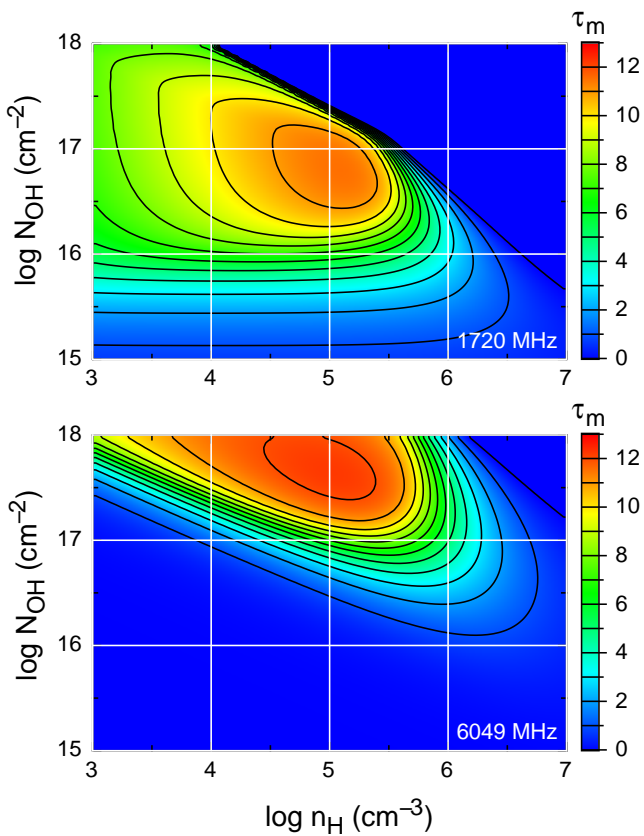}
  \caption{Maser optical depth as a function of $\nH$
and $N_\mathrm{OH}$ for T = 50\,K for the 1720 MHz (top) and 6049
MHz (bottom) lines of OH.}\label{fig:contours}
\end{figure}

Figs. \ref{fig:tau} and \ref{fig:contours} show results for $T=50$\,K, assuming small
velocity gradients within the slab, and no far-infrared
radiation from dust.  These conditions are broadly consistent with
warm gas that is cooling off behind a shock wave driven into
a molecular cloud by an adjacent supernova remnant.  Fig.
\ref{fig:tau} shows the maser optical depth through the slab
for the 1720 MHz satellite line in the ground rotational state and its
analogues in the first (6049 MHz) and second (4765 MHz) excited
rotational states.  At low OH column densities the 1720 MHz line is
inverted, peaking in the range $10^{16}$--$10^{17}\persqcm$, as expected.  At about
$10^{17}\persqcm$ the optical depth of the slab to the 1720 MHz line
suppresses the inversion.  Meanwhile the inversion in the 6049\,MHz
line grows, peaking at $N_\mathrm{OH}\approx 3\times 10^{17}\persqcm$.
The 4765\,MHz line peaks at $N_\mathrm{OH}\sim 10^{18}\persqcm$.
  The effect
of varying $\nH$ and $N_\mathrm{OH}$ is explored in Fig.\
\ref{fig:contours}.  As expected, masing in the 1720\,MHz line is
strongest for $\nH\sim 10^5\percc$ and $N_\mathrm{OH}\sim
10^{16.5}\persqcm$.  The inversion of the 6049\,MHz line requires
column densities in excess of $10^{17}\persqcm$, while inversion at
4765\,MHz (not shown) requires an OH column a few times higher still.

\section[]{Observations and data reduction}
\label{sec:observations}

The observations were made 2006 September 6--7 with the Parkes 64-m
radio telescope and the Methanol Multibeam receiver.  The
seven-beam receiver covers a frequency range of 6.0--6.7 GHz, producing
two 300 MHz IF bands in dual circular polarisation.  The beamsize is
3.2 arcmin at 6.7 GHz and 3.4 arcmin at 6.0 GHz.  At the frequencies
observed, 6049, 6035 and 6030~MHz, this was assumed to be 3.4 arcmin.  The beams are
hexagonally configured with a beam separation on the sky of 6.462 arcmin.  The feed was rotated 19.1 degrees with respect to the
scan direction, such that the beams trace a comb of equally spaced
paths, each 0.55 arcmin apart.  The chosen correlator configuration allowed an 8~MHz bandwidth with 2048 channels for each IF, giving a channel width
of 0.195~km~s$^{-1}$ at the observed frequencies.  The pointing accuracy of the Parkes Telescope is better than 20
arcsecs. 

The 6030 and 6035~MHz lines were observed in one IF with 6049~MHz being observed
in the other.  The 6030 and 6035~MHz fields were recentred such
that the velocity centre was midway between the two frequencies and
adjusted for the velocity of the remnant and the observatory in the
Local Standard of Rest (LSR).  If the velocity of the remnant was unknown, the frequency centre was
estimated from supernova remnants nearby in RA and Dec.  As the
bandwidth of the second IF should be
wide enough to include 6049 MHz emission at galactic velocities, only the extragalactic fields were
recentered to account for their velocity at 6049~MHz. 

Scans in right ascension and declination were conducted over the
supernova remnants at a scan rate of 0.1 deg/min, with a total
observing time in one scanning direction of
$\sim$100 minutes per square degree, which corresponds to 10 scans.
Based on the sizes of the remnants, scans were conducted over either
0.25, 0.5, 0.75 or 1.0 degree areas to completely cover the area of the
remnant.   The average rms noise ranges from 0.1 to 0.2 Jy/beam for an object
scanned in both RA and Dec and is $\sqrt{2}$ higher if scanned in
only one direction.

The primary flux calibrator observed was 1934$-$638.  It has a positional
accuracy of 0.02~arcsec.  Each beam was scanned in RA then Dec through
1934$-$638 and the power of a noise calibrator was compared with the known flux
density of 1934$-$638, which is 4.4~Jy at 6035~MHz.  The calibrator was observed twice, but it was
raining heavily during one of the observations so only one of the
calibrator scans was used.  The weather was cloudy or raining during
all the observations and this
adds to the uncertainty in the flux calibration.

Thirty-five supernova remnants, a star-forming region and four fields in the Large and Small Magellanic
Clouds were observed (Table \ref{Tab:obs}).  The remnants were chosen such that the majority
were known or thought to be interacting with molecular clouds.  The
star-forming region has a known 6035~MHz OH maser and was observed to
confirm that the system was working.  Eighteen of
the galactic supernova remnants chosen have detected 1720~MHz OH
masers
(\citealt{1994ApJ...424L.111F}; \citealt*{1995Sci...270.1801Y}; \citealt{1996AJ....111.1651F};
\citealt{1996ApJ...466L..25Y}; \citealt{1997ApJ...489..143C};
\citealt{1997AJ....114.2058G}; \citealt{1998AJ....116.1323K}; \citealt{1999ApJ...527..172Y}).
The star-forming region 30~Doradus in the Large Magellanic Cloud (LMC)
has a detected 1720~MHz maser and the supernova remnant N49, also in
the LMC, is thought to have a 1720~MHz maser
(\citealt{2004AJ....128..700B}; \citealt{2005AJ....129..805R}).
However, as it is predicted that 1720~MHz OH masers are unlikely to
be produced when the column density is high enough to produce 6049~MHz
OH masers, half of the supernova remnants that were observed
have no 1720~MHz masers, but other signatures indicating an interaction with a molecular cloud.  Eleven of the remnants surveyed are a part of the
mixed-morphology class of supernova remnants with an additional nine
possibly being members of this class
\citep{1998ApJ...503L.167R,2003ApJ...585..319Y}.  Other signatures suggestive of an interaction with a molecular cloud
such as TeV $\gamma$-ray emission
\citep{2003PASJ...55L..61F}, infrared emission suggesting molecular
shocks (e.g. \citealt{2006AJ....131.1479R}), or the presence of shocked CO or nearby
CO clouds (e.g. \citealt{1999AJ....118..930D}), were used to identify prospective
galactic supernova remnants.  

Two fields in the Small Magellanic Cloud (SMC) were observed when no
galactic supernova remnants were above the telescope horizon.  The fields were
chosen to contain as many supernova remnants as possible - nine in the
first field and eight in the second \citep{2008A&A...485...63F}. 

The data were flux-calibrated, continuum subtracted and gridded using
the programs Livedata and Gridzilla \citep{2001MNRAS.322..486B}.  Livedata is a program
specifically designed for data taken with the Parkes Multibeam
receiver.  It removes the bandpass, calibrates the spectra, applies
doppler tracking, and smoothes and removes the baselines of the spectrum.  The
bandpass calibration uses robust statistical estimators, such as
median estimators, which can reject radio frequency interference without human intervention and
is tailored to identify compact emission regions.  Gridzilla converts the individual
spectra produced by Livedata into position-position-velocity cubes by
placing the spectra on a regular grid.  The gridding process
determines which spectra will contribute to an individual pixel,
rejects those spectra which appear to contain corrupt data and
ascribes a statistical weight to each remaining spectrum.  The value of the pixel
is then calculated based on the input data and weights.  As with
Livedata, the gridding process is optimised for point sources and uses
the median statistic to produce more robust results. A scale of 1 arcmin per pixel was used.

The data cubes were
first searched for any obvious maser emission by eye and then searched using the program Duchamp\footnote{http://www.atnf.csiro.au/people/Matthew.Whiting/Duchamp/}.  Duchamp is designed to search for small, isolated
objects within spectral-line data cubes.  The program searches for
connected voxels that are above some flux threshold, with user-defined
limits on size and no assumption made as to the shape of the
detections.

The cube can be reconstructed using the \textit{\`{a} trous} wavelet
procedure, which effectively reduces much of the noise in the image,
thereby allowing the cube to be searched to fainter levels while
reducing the number of spurious detections.  The reconstruction is
conducted by discretely convolving the input array with a B$_3$-spline
filter function, determining wavelet coefficients by calculating the
difference between the convolved array and the input array and if
these coefficients are above the requested threshold at a given point, they are
added to the reconstructed array.  The separation between the filter
coefficients is then doubled and the first step is performed again
with the convolved array as the input array.  This process is
continued until the maximum number of scales required is reached and
then the final smoothed array is added to the reconstructed array to
provide the `DC offset'.

The 2-dimensional raster-scanning algorithm of \citet{1980CompJ..23..262L} is used to
search the cube for detections, one channel map at a time.  The middle
of the noise distribution and the width of the distribution is used in
calculating the threshold.  The threshold is determined using the
False Discovery Rate technique \citep{2001AJ....122.3492M,2002AJ....123.1086H}.  This is given by the
number of false detections divided by the total number of detections
and is set by the user.  In practice, the final list of detected
objects that are false positives will be much smaller than that
requested by the user as, due to merging and rejection of the
individual pixels, there is not a
direct connection between this value and the fraction of detections
that will be false positives.  

The repeated detections in the list produced are then combined in an
algorithm that matches objects judged to be `close'.  The threshold
distances apart for two objects to be merged spatially and in the
velocity axis are set by the user.  The spatial requirement can be set
such that there must be a pair of pixels that are adjacent.  Once the
detections are merged, they can be grown to some secondary
threshold and then are sent through the merging algorithm again.  To
be accepted as detections, the objects must span both a minimum number
of velocity channels and a minimum number of spatial pixels.

The \textit{\`{a} trous} reconstruction was used with a false
discovery rate of 10
percent.  The detections were restricted
to being adjacent in space and velocity and a minimum of two velocity
channels wide and 12 spatial pixels in size.  

\begin{table*}
\begin{minipage}{177mm}
\caption{Observed sources}
\label{Tab:obs}
\begin{tabular}{@{}lcclcccccl@{}}
\hline
Source	&	RA	&	Dec	&	Centre	&	Size	&	Area	&	No. of	&	\multicolumn{2}{c}{RMS noise}			&	Notes	\\
	&	(J2000)	&	(J2000)	&	V$_{\mathrm{LSR}}$	&	(arcmin)	&	(deg)	&	scans	&	\multicolumn{2}{c}{(Jy/beam)}			&		\\
	&	(h m s)	&	(\degr\, \arcmin\, \arcsec)	&	(km s$^{-1}$)	&		&		&	(RA/Dec)	&	6035	&	6049	&		\\
\hline																			
N49	&	05 26 01	&	$-$66 05 00	&	+278	&	$1.5\times 1.5$	&	0.25	&	2/2	&	0.11	&	0.13	&	m$^1$	\\
30 Doradus	&	05 38 42	&	$-$69 06 00	&	+242	&	$4.5\times 0.6$	&	0.25	&	2/2	&	0.11	&	0.13	&	S$^2$, m$^3$	\\
G240.315+0.071	&	07 44 52	&	$-$24 08 00	&	+63.6	&		&	0.25	&	1/1	&	0.15	&	0.17	&	S$^4$	\\
G272.2$-$3.2	&	09 06 50	&	$-$52 07 00	&	$-$77	&	15	&	0.50	&	1/1	&	0.11	&	0.13	&	MM$^5$	\\
G284.3$-$1.8	&	10 18 15	&	$-$59 00 00	&	$-$10	&	24	&	0.50	&	1/1	&	0.11	&	0.13	&	I$^6$	\\
IC 443 (G189.1+3.0)	&	06 17 00	&	+22 34 00	&	$-$4	&	45	&	0.75	&	1/1	&	0.11	&	0.12	&	m$^7$, MM$^8$, F, H	\\
G286.5$-$1.2	&	10 35 50 	&	$-$59 42 00	&	$-$40	&	$26\times6$	&	0.50	&	1/1	&	0.11	&	0.13	&		\\
G289.7$-$0.3	&	11 01 15	&	$-$60 18 00	&	$-$35	&	$18\times14$	&	0.50	&	1/1	&	0.11	&	0.13	&		\\
G290.1$-$0.8 (MSH11$-$61A)	&	11 03 05	&	$-$60 56 00	&	+15	&	$19\times14$	&	0.50	&	1/1	&	0.12	&	0.13	&	MM$^8$	\\
G292.0+1.8 (MSH 11$-$54)	&	11 24 36	&	$-$59 16 00	&	$-$36	&	$12\times8$	&	0.50	&	1/1	&	0.11	&	0.13	&		\\
Kes17 (G304.6+0.1)	&	13 05 59	&	$-$62 42 00	&	$-$38	&	8	&	0.25	&	1/1	&	0.16	&	0.19	&	I?$^9$	\\
G327.1$-$1.1	&	15 54 25	&	$-$55 09 00	&	$-$38	&	18	&	0.50	&	1/1	&	0.12	&	0.14	&		\\
Kes 27 (G327.4+0.4)	&	15 48 20	&	$-$53 49 00	&	$-$38	&	21	&	0.50	&	P/1	&	0.12	&	0.14	&	MM$^8$	\\
RCW 103 (G332.4$-$0.4)	&	16 17 33	&	$-$51 02 00	&	$-$48	&	10	&	0.25	&	1/1	&	0.13	&	0.18	&	I$^{10}$	\\
CTB 33 (G337.0$-$0.1)	&	16 35 57	&	$-$47 36 00	&	$-$69	&	1.5	&	0.25	&	1/1	&	0.16	&	0.18	&	m$^{11}$	\\
Kes 41 (G337.8$-$0.1)	&	16 39 01	&	$-$46 59 00	&	$-$45	&	$9\times6$	&	0.25	&	1/1	&	0.16	&	0.19	&	m$^{12}$, MM?$^{13}$	\\
G346.6$-$0.2	&	17 10 19	&	$-$40 11 00	&	$-$76	&	8	&	0.25	&	1/1	&	0.16	&	0.18	&	m$^{12}$, MM?$^{13}$	\\
G347.3$-$0.5	&	17 13 50	&	$-$39 45 00	&	$-$6	&	$65\times55$	&	1.00	&	0/1	&	0.14	&	0.16	&	I$^{14}$	\\
CTB37A (G348.5+0.1)	&	17 14 06	&	$-$38 32 00	&	$-$42	&	15	&	0.50	&	1/1	&	0.12	&	0.14	&	m$^{11}$, H	\\
G349.7+0.2	&	17 17 15	&	$-$38 04 00	&	+15	&	$9\times6$	&	0.25	&	1/1	&	0.17	&	0.19	&	m$^{11}$, MM?$^{13}$, H	\\
G357.7+0.3	&	17 38 35	&	$-$30 44 00	&	$-$35	&	24	&	0.50	&	0/1	&	0.20	&	0.21	&	m$^{15}$, MM?$^{13}$, H	\\
G357.7$-$0.1 (MSH 17$-$39)	&	17 40 29	&	$-$30 58 00	&	$-$12	&	$8\times3$	&	0.25	&	1/1	&	0.17	&	0.20	&	m$^{11}$, MM?$^{13}$, H	\\
G359.1$-$0.5	&	17 45 30	&	$-$29 57 00	&	$-$5	&	24	&	0.50	&	1/1	&	0.15	&	0.17	&	m$^{16}$, MM$^{13}$, H	\\
Sgr A East (G0.0+0.0)	&	17 45 44	&	$-$29 00 00	&	+58	&	$3.5\times2.5$	&	0.25	&	1/1	&	0.19	&	0.21	&	m$^{17}$, MM$^{13}$, F	\\
Sgr D (G1.1$-$0.1)	&	17 48 30	&	$-$28 09 00	&	$-$1	&	8	&	0.25	&	1/1	&	0.16	&	0.19	&	m$^{15}$, MM?$^{13}$, F	\\
G1.4$-$0.1	&	17 49 39	&	$-$27 46 00	&	$-$2	&	10	&	0.25	&	1/1	&	0.16	&	0.18	&	m$^{15}$, F	\\
W 28 (G6.4$-$0.1)	&	18 00 30	&	$-$23 26 00	&	+9	&	48	&	1.00	&	1/1	&	0.11	&	0.12	&	m$^{18}$, MM$^8$, F, H	\\
G11.2$-$0.3	&	18 11 27	&	$-$19 25 00	&	+65	&	4	&	0.25	&	1/1	&	0.16	&	0.18	&		\\
G16.7+0.1	&	18 20 56	&	$-$14 20 00	&	+20	&	4	&	0.25	&	0/1	&	0.22	&	0.25	&	m$^{19}$, MM?$^{13}$, F, H	\\
Kes 67 (G18.8+0.3)	&	18 23 58	&	$-$12 23 00	&	+18	&	$17\times11$	&	0.50	&	1/1	&	0.12	&	0.14	&	I$^{20}$	\\
Kes 69 (G21.8$-$0.6)	&	18 32 45	&	$-$10 08 00	&	+69	&	20	&	0.50	&	1/1	&	0.12	&	0.13	&	m$^{19}$, MM?$^{13}$, F, H	\\
G22.7$-$0.2	&	18 33 15	&	$-$09 13 00	&	+81	&	26	&	0.50	&	P/1	&	0.15	&	0.17	&	I$^9$	\\
3C 391 (G31.9+0.0)	&	18 49 25	&	$-$00 55 00	&	+110	&	$7\times5$	&	0.25	&	P/1	&	0.16	&	0.18	&	m$^{11}$, MM$^8$, F, H	\\
Kes 79 (G33.6+0.1)	&	18 52 48	&	$-$00 41 00	&	+110	&	10	&	0.25	&	1/1	&	0.16	&	0.18	&	I$^{21}$, MM?$^8$	\\
W 44 (G34.7$-$0.4)	&	18 56 00	&	+01 22 00	&	+45	&	$35\times27$	&	0.75	&	0/1	&	0.17	&	0.17	&	m$^7$, MM$^8$, F, H	\\
3C 397 (G41.1$-$0.3)	&	19 07 34	&	+07 08 00	&	+40	&	$4.5\times2.5$	&	0.25	&	1/1	&	0.15	&	0.17	&	I?$^{22}$	\\
W 49B (G43.3$-$0.2)	&	19 11 08	&	+09 06 00	&	+48	&	$4\times3$	&	0.25	&	1/1	&	0.16	&	0.18	&	I?$^{23}$	\\
W 51C (G49.1$-$00.1)	&	19 23 19	&	+14 09 00	&	+70	&	$25\times19$	&	0.50	&	0/1	&	0.15	&	0.17	&	m$^{19}$, MM$^8$, F, H	\\
3C 400.2 (G53.6$-$2.2)	&	19 38 50	&	+17 14 00	&	+29	&	$33\times28$	&	0.75	&	0/1	&	0.14	&	0.16	&	MM$^8$	\\
SMC field 1	&	01 01 50	&	$-$71 58 00	&	+150	&	$60\times60$	&	1.00	&	1/1	&	0.10	&	0.11	&	9 SNRs$^{24}$	\\
SMC field 2	&	00 51 00	&	$-$73 00 00	&	+150	&	$60\times60$	&	1.00	&	1/1	&	0.11	&	0.12	&	8 SNRs$^{24}$	\\
\hline
\end{tabular}
\\
P: Partial scan of area in this direction, MM: mixed-morphology, S: star forming
region, I: other signatures indicating interaction with molecular cloud, m: source with
1720 MHz masers, F: SNRs with pointings observed at 6049~MHz by
\citet{2007ApJ...670L.117F}, H: SNRs with pointings observed at 1612,
1665, 1667 and 1720~MHz by \citet{2006ApJ...652.1288H}

References -- (1) \citealt{2004AJ....128..700B} (2)
\citealt{1987ApJ...323L..65W} (3) \citealt{2005AJ....129..805R} (4)
\citealt*{1996A&AS..115...81B} (5) \citealt*{1994A&A...286L..35G} (6)
\citealt{1986ApJ...309..667R} (7) \citealt{1997ApJ...489..143C} (8)
\citealt{1998ApJ...503L.167R} (9) \citealt{2006AJ....131.1479R} (10)
\citealt*{1989A&A...214..307O} (11) \citealt{1996AJ....111.1651F} (12)
\citealt{1998AJ....116.1323K} (13) \citealt{2003ApJ...585..319Y} (14)
\citealt{2003PASJ...55L..61F} (15) \citealt{1999ApJ...527..172Y} (16)
\citealt{1995Sci...270.1801Y} (17) \citealt{1996ApJ...466L..25Y} (18)
\citealt{1994ApJ...424L.111F} (19) \citealt{1997AJ....114.2058G} (20)
\citealt{1999AJ....118..930D} (21) \citealt{1992MNRAS.254..686G} (22)
\citealt{2005ApJ...618..321S} (23) \citealt{2007ApJ...654..938K} (24) \citealt{2008A&A...485...63F}
\end{minipage}
\end{table*}

\section{Results}
\label{sec:results}

\begin{figure*}
\begin{minipage}{155mm}
 \includegraphics[width=155mm]{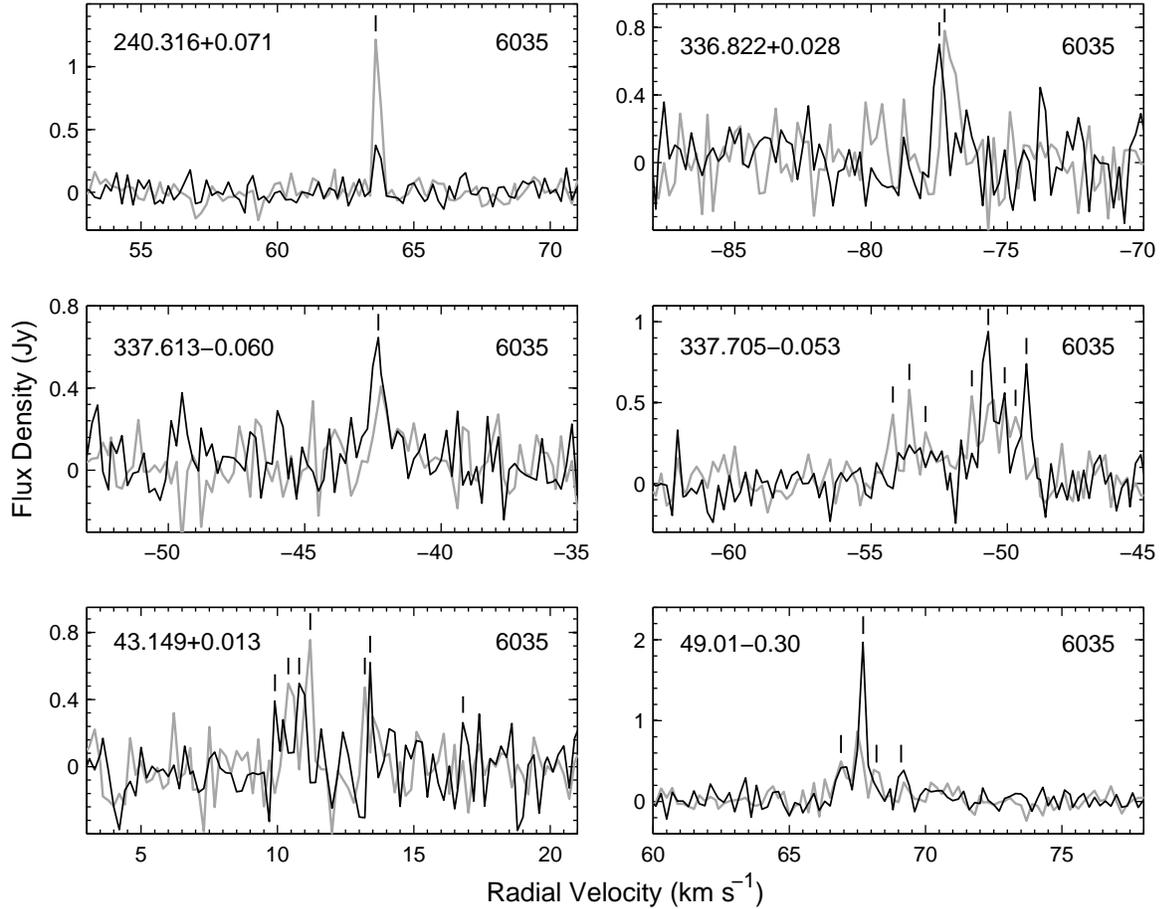}
                                
  \caption{Spectra of OH 6035~MHZ maser emission.  The channel width is 0.195~km~s$^{-1}$ and
  the beamsize is 3.4 arcmin.  The date of observation is 2006
  September 6--7.  The grey line denotes left-hand circular polarisation
  and the black line denotes right-hand circular polarisation.  Maser
  peaks are marked.  Note that the total intensity of each source is the sum (not the average) of the
  two polarisations.}\label{fig:6035}
\end{minipage}
\end{figure*}

\begin{figure}
 \includegraphics{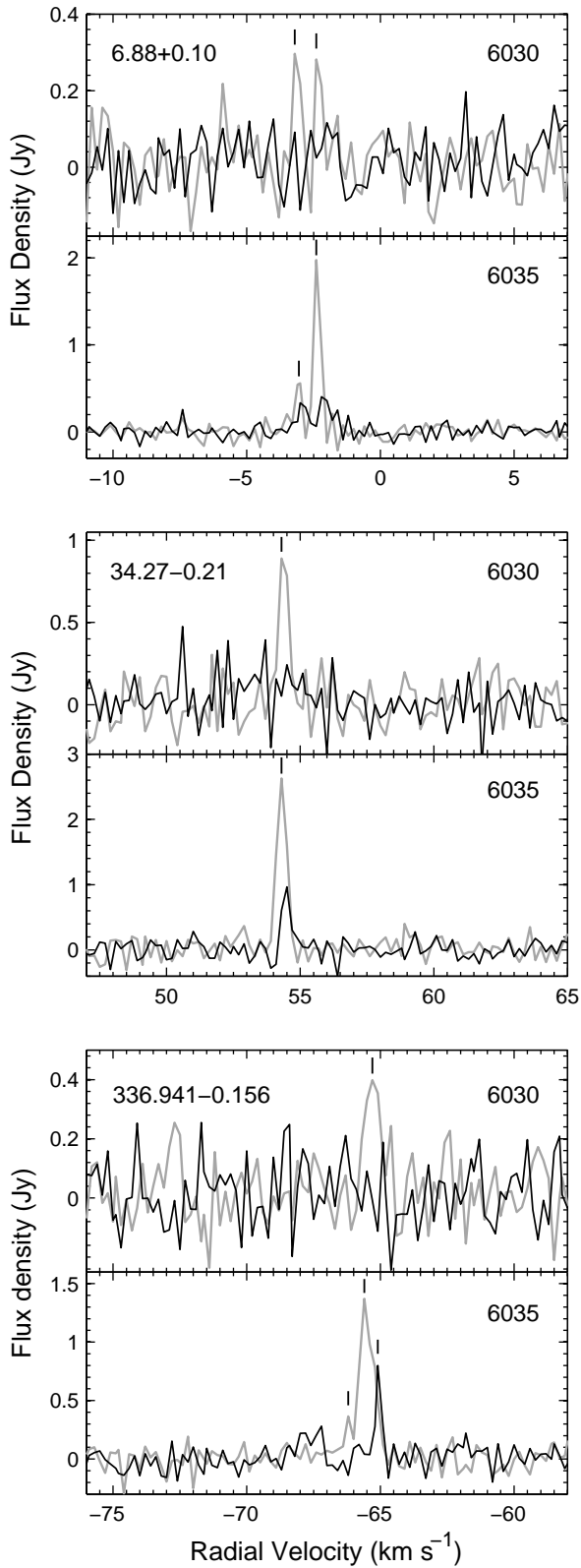}
                                
  \caption[width=84mm]{Spectra of 6030~MHz OH masers with associated
    6035~MHz maser emission.  The channel width is 0.195~km~s$^{-1}$ and
  the beamsize is 3.4 arcmin.  The date of observation is 2006
  September 6--7.  The grey line denotes left-hand circular polarisation
  and the black line denotes right-hand circular polarisation.  Maser
  peaks are marked.  Note that the total intensity of each source is the sum (not the average) of the
  two polarisations.}\label{fig:60306035}
\end{figure}

No 6049~MHz OH maser emission was detected to a 3-$\sigma$ level of
0.3 to 0.6~Jy/beam, with the noise limits for each SNR listed in
Table~\ref{Tab:obs}.  These limits are for the average of the
two circular polarisations.  However, eleven masers at 6035 and three at 6030~MHz not associated with the supernova
remnants were detected.  These masers are likely to be associated with
star-forming regions.  Fig. \ref{fig:6035} shows the spectra of the
six sources with only
6035~MHz emission detected.  The 6035~MHz source 43.149+0.013
contained in this figure is a confused source with spectral features
from three masers: 43.149+0.013, 43.165+0.013 and 43.165$-$0.028.   Spectra of the three detected 6030~MHz sources
aligned in velocity with their associated 6035~MHz sources are shown in
Fig. \ref{fig:60306035}, two of which are new detections.          

 \begin{table*}
 \centering
 \begin{minipage}{177mm}
\caption{Detected 6035- and 6030- MHz OH masers.  The
  velocity and flux of the brightest peaks are given if
  multiple peaks are observed.}
\label{Tab:masers}
\begin{tabular}{@{}llllcccccccc@{}}
\hline
	&		&		&		&	\multicolumn{2}{c}{6035 MHz}			&	\multicolumn{2}{c}{6035 MHz}			&	\multicolumn{2}{c}{6030 MHz}			&	Magnetic	&	Notes	\\
	&	OH maser	&	RA (2000)	&	Dec (2000)	&	\multicolumn{2}{c}{Velocity Peak}			&	\multicolumn{2}{c}{Peak Flux}			&	\multicolumn{2}{c}{Peak Flux}			&	Field	&		\\
Field	&	(l b)	&		&		&	\multicolumn{2}{c}{(km~s$^{-1}$)}			&	\multicolumn{2}{c}{(Jy)}			&	\multicolumn{2}{c}{(Jy)}			&	(mG)	&		\\
	&	($\degr$ $\degr$)	&	(h m s)	&	($\degr$ $\arcmin$ $\arcsec$)	&	L	&	R	&	L	&	R	&	L	&	R	&		&		\\
\hline																							
G240.315+0.071	&	240.316+0.071	&	07 44 51.97	&	$-$24 07 42.3	&	+63.6	&	+63.6	&	1.22	&	0.37	&	-	&	-	&	-	&		\\
CTB 33	&	336.822+0.028	&	16 43 38.29	&	$-$47 36 32.9	&	$-$77.3	&	$-$77.5	&	0.78	&	0.70	&	-	&	-	&	$-$3.6	&	E	\\
CTB 33	&	336.941$-$0.156	&	16 35 55.20	&	$-$47 38 45.4	&	$-$65.6	&	$-$65.1	&	1.37	&	0.80	&	0.40	&	-	&	+10.4	&		\\
Kes 41	&	337.613$-$0.060	&	16 38 09.54	&	$-$47 04 59.9	&	$-$42.2	&	$-$42.4	&	0.41	&	1.35	&	-	&	-	&	$-$3.6	&		\\
Kes 41	&	337.705$-$0.053	&	16 38 29.67	&	$-$47 00 35.8	&	$-$53.6	&	$-$50.7	&	0.58	&	0.94	&	-	&	-	&	-	&		\\
W28	&	6.88+0.10	&	18 00 49	&	$-$22 57 26	&	$-$2.4	&	$-$2.2	&	1.98	&	0.40	&	0.30	&	-	&	+3.6	&	N	\\
W44	&	34.27$-$0.21	&	18 54 37	&	+01 05 29	&	+54.3	&	+54.5	&	2.63	&	0.97	&	0.89	&	0.48	&	+3.6	&N		\\
W49B	&	43.149+0.013	&	19 10 11.05	&	+09 05 22.1	&	+11.2	&	+10.8	&	0.76	&	0.46	&	-	&	-	&	text	&	E,C	\\
W51C	&	49.01$-$0.30	&	19 22 29	&	+14 07 24	&	+67.5	&	+67.7	&	0.87	&	1.97	&	-	&	-	&	+3.6	&		\\
\hline
\end{tabular}
\\
N: newly detected maser, E: maser on spatial edge of field, C: confused source.
\end{minipage}
\end{table*}

Table \ref{Tab:masers} contains the list of identified masers.  The
field in which the maser was located is listed in column 1 and the galactic
longitude and latitude is used as a source name, following the usual
practice, and is shown in column 2.  The coordinates for each source
are given in columns 3, 4 (equinox
J2000) and refer to the position
of the strongest feature (taken from higher resolution observations
for the previously known masers).  The
positional uncertainty for the positions derived from the current
observations is 15 arcsec.  The velocity of the strongest left- and right-hand
circular polarisation components (LHCP and RHCP) at 6035~MHz are given in columns
5 and 6 and the peak flux at 6035~MHz is given in columns 7 and 8.
The peak flux at 6030~MHz for both polarisations are listed in columns
9 and 10 and the estimated magnetic field is given in column 11, where a
field directed away from us is positive and is indicated by
the RHCP feature at a more positive velocity than the LHCP feature.
Column 12 gives notes about the maser detection.  As per convention, spectra are scaled such that the total
intensity is the sum, not the average, of the two circular
polarisations. The continuum levels in the spectra varied from
0.01--0.05~Jy.  The spectra have had this continuum offset removed and
the fluxes were measured with respect to the continuum level.    

As left- and right-hand circular polarisations were observed, Zeeman
pairs can be identified.  A 1~mG line-of-sight magnetic field component produces splittings
equivalent to 0.079~km~s$^{-1}$ and 0.056~km~s$^{-1}$
in the 6030 and 6035~MHz transitions respectively
\citep{1974IAUS...60..275D}.  Thus a difference of one channel width
between the left and right polarisations at 6035~MHz corresponds to a
line-of-sight magnetic field of 3.6~mG.  The observed splittings were
one to three channels so the magnetic fields are rough estimates, with
an estimated uncertainty of
approximately 2~mG (see Table \ref{Tab:masers}).

Individual maser sources are discussed below.

\textit{240.316+0.071}.  (Field: G240.315+0.071, spectrum in Fig. \ref{fig:6035})  This source was first observed by \citet{1995MNRAS.273..328C}
and has remained stable with the intensity of the strongest feature
at approximately 2~Jy in LHCP through
previous observations \citep{2003MNRAS.341..551C}.  Our new
observations show that the strongest features of both the LHCP and the RHCP
have halved in intensity and a secondary peak at
approximately +62~km~s$^{-1}$ is no longer visible in the spectrum for
either polarisations.  In agreement with previous observations, no Zeeman splitting was detected.

\textit{336.822+0.028}.  (Field: CTB 33, spectrum in Fig. \ref{fig:6035})  This source was on the spatial edge of the field in
which it was located.  The spectrum has not changed much from
\citet{1995MNRAS.273..328C}, the only real difference is a slight increase in
intensity of the single LHCP and RHCP feature.  Assuming that the
features are a Zeeman pair, with the LHCP feature at a larger velocity
than the RHCP by 0.2~km~s$^{-1}$, a magnetic field of $-3.6$~mG is
determined.  Within the spectral resolution of our observations, this
is in agreement with the value of $-5$~mG determined by \citet{1995MNRAS.273..328C}.

\textit{336.941$-$0.156}.  (Field: CTB 33, spectrum in Fig. \ref{fig:60306035})  The 6035~MHz emission was discovered by
\citet{2001MNRAS.326..805C} and the 6030~MHz by \citet{2003MNRAS.341..551C}.  There are
significant differences between the previous spectra and the new
observations.  There was minimal RHCP
6035~MHz emission detected in 2001, in comparison to a strong feature
detected in the current observations.  As the RHCP feature is at a
larger velocity than the LHCP by 0.6~km~s$^{-1}$, a magnetic field of
+10.4~mG is calculated.  With the lower spatial resolution of the
current observations, it is not possible to determine whether the
6030~MHz LHCP maser emission contains the two features present in the
2001 observations.  The RHCP 6030~MHz emission was not strong enough to be
detected.     

\textit{336.983$-$0.183}.  (Field: CTB 33)  This weak source discovered by \citet{2001MNRAS.326..805C} was
below our detection limit.

\textit{337.613$-$0.060}.  (Field: Kes 41, spectrum in Fig. \ref{fig:6035})  The 2001 observations showed marked changes
from the 1994 observations, with the strongest feature at $-42$~km~s$^{-1}$ halving in intensity and a new LHCP feature at $-41.1$~km~s$^{-1}$ of 1 Jy becoming the strongest feature \citep{2003MNRAS.341..551C}.  Our 2006
observations also show change, with the RHCP feature at
$-42.4$~km~s$^{-1}$ now being
the strongest feature and no obvious feature at
$-41.1$~km~s$^{-1}$.  The magnetic field is approximately
$-3.6$~mG, which is in agreement with that derived from
the 1994 observations and from an
associated 1665~MHz maser \citep{1995MNRAS.273..328C}.

\textit{337.705$-$0.053}.  (Field: Kes 41, spectrum in Fig. \ref{fig:6035})  The new observations are similar to the
spectrum from \citet{1995MNRAS.273..328C}, the major difference being that the RHCP polarisation between $-55$ and
$-52$~km~s$^{-1}$ is much lower.  The RHCP is
consistently at a more positive velocity than the LHCP, indicating a
Zeeman pattern and in agreement with the previous observations, but as the features are blended together and the
spectral resolution is not very high, no magnetic field could be estimated.   

\textit{6.88+0.10}.  (Field: W28, spectrum in Fig. \ref{fig:60306035})  This 6030 and strong 6035~MHz source is a new
maser with RHCP possibly at a larger velocity than LHCP at 6035~MHz by
0.2~km~s$^{-1}$, corresponding to a magnetic field of +3.6 mG, assuming that the
features form a Zeeman pair.

\textit{22.435$-$0.169}. (Field: G22.7-0.2)   This source discovered by \citet{1995MNRAS.273..328C} was below our detection limit.  

\textit{34.27$-$0.21}.  (Field: W44, spectrum in Fig. \ref{fig:60306035})  This new 6030 and strong 6035~MHz maser may be
what was characterised a `possible' 6035 MHz detection at coordinates 34.3$-$0.15 and velocity
56.0~km~s$^{-1}$ by \citet{1984A&A...131...45G}.  The 6035~MHz maser
is characterised by a single feature in both polarisations and the
RHCP is at a larger velocity than the LHCP by 0.2~km~s$^{-1}$, giving
a magnetic field of +3.6~mG.  The LHCP 6030~MHz emission is comparable
to the intensity of the RHCP 6035~MHz emission.  

\textit{43.149+0.013}, \textit{43.165+0.013} and
\textit{43.165$-$0.028}.  (Field: W49B, spectrum in Fig. \ref{fig:6035})  These three masers are located within W49
North and South.  43.149+0.013 and 43.165+0.013 lie in the northern
complex and are separated by
nearly 60~arcsec \citep{2001MNRAS.326..805C} while 43.165$-$0.028 is nearly 3~arcmin away
from the other two in the southern complex.  Emission from these three
masers was confused in the current observations as the beam size was
3.4~arcmin.  The emission was also on the edge of the field.  The
maser emission is weaker than detected by \citet{1995MNRAS.273..328C} and the
features at +17~km~s$^{-1}$ are barely visible, but this may be because the emission is on the very edge of the field.  A
magnetic field is not calculated due to the above confusion, but the
larger velocity of the LHCP with respect to the RHCP for the features at +10~km~s$^{-1}$ to
+12~km~s$^{-1}$ agrees with previous observations.

\textit{49.01$-$0.30}.  (Field: W51C, spectrum in Fig. \ref{fig:6035})  This 6035 and 6030~MHz maser was detected by
\citet{1997A&A...325..255B} at the position 48.99$-$0.30.  6030~MHz was
not detected in our current observations and may be below the
detection limit.  The 6035~MHz spectrum is similar to the previous
observation, with the major difference being a large reduction in the
brightness of the feature at +69~km~s$^{-1}$.  The RHCP at 6035~MHz
is at a larger velocity than the LHCP by 0.2~km~s$^{-1}$, giving
a possible magnetic field of +3.6~mG.    

The 6035~MHz maser at 49.01$-$0.30 is coincident with a compact H~\small{\textsc{II}} \normalsize region
G49.0$-$0.30 and an IRAS source which is believed to be an ultracompact
H~\small{\textsc{II}} \normalsize region \citep{1999ApJ...518..760K}.  The radio
recombination line velocity of G49.0$-$0.30 is between
+64--66~km~s$^{-1}$ \citep*{1979A&A....75..365P}, which is consistent with
the velocity of the maser.  

While not all of the masers discussed above have known associated
H~\small{\textsc{II}} \normalsize regions or ultracompact H~\small{\textsc{II}}
\normalsize regions, it is expected that they are all associated with
star-forming regions.  The variability seen in the masers is
expected.  \citet{1995MNRAS.273..328C}, for example, discussed the
variability of 6035~MHz OH masers in star-forming regions.  They found that while there seemed
to be no perceptible changes over an interval of a few days (although
shorter time-scale variations had been reported previously, e.g. the transient
appearance and disappearance of a feature in 351.420+0.64 in less than
12 hours by \citealt{1972ApJ...177...59Z}), some
changes could be seen on a 6-month interval, and many changes could be
seen over a 20-year period, including more than 50 percent changes in intensity
in some features.  Our observations show that some of the masers have
changed in intensity from the 2001 to the 2006
observations, with some strong features halving and some features
disappearing completely.

\section{Discussion}
\label{discussion}
\citet{2007ApJ...670L.117F} observed pointings at 6016, 6030, 6035 and
6049~MHz towards 14 supernova
remnants using the Effelsberg 100-m telescope, 10 of which were
covered in our observations (see Table \ref{Tab:obs} column 10), during 2007 July.
The sensitivity of their observations was an order of magnitude
better, with a 5-$\sigma$ rms of 30~mJy.  No
masers were detected towards any of the supernova remnants observed down to
this level, in agreement with our observations.

Eleven 6035~MHz OH masers and three OH 6030~MHz masers
were detected in our study.  These masers are generally associated with star-forming regions
rather than supernova remnants.  1665~MHz OH masers are usually the strongest OH transition in
star-forming regions, but weaker masers at other OH transitions, such as
6030 and 6035 MHz are also often detected.  Models of masers in
star-forming regions predict these transitions.  For example,
\citet*{1991MNRAS.252...30G} and \citet*{1992A&A...262..555G} found
that bright 6035~MHz masers are formed in warm dense regions where
collisions play an important role in creating the inversion.  The multiple maser spots
at each site arise in the dusty molecular envelope around a massive star
in its early stages of formation.  The maser spots are valuable probes that can reveal the
physical parameters of the environment where the massive star is
forming, at a stage before it has been disrupted by the embedded star,
and before the visible light can emerge through the dust.  The maser
phase occurs at the same time as the development of an ultracompact
H~\small{\textsc{II}}\normalsize\,region around the star, but fade
rapidly when the H~\small{\textsc{II}}\normalsize\,region has expanded to a size greater than
$\sim 5000$~AU \citep{2000ApJ...530..371F}.

\citet{2008MNRAS.385..948G} have conducted a survey for 6035~MHz OH
and 6668~MHz methanol masers in the Magellanic Clouds.  With a
3-$\sigma$ limit of 0.39~Jy, they did not
detect any 6035~MHz masers in the SMC, which is in agreement with our
lack of detections for the two fields in the SMC.          

No OH 6049~MHz maser emission was detected in the current work above our 3-$\sigma$
levels, which range from 0.3
to 0.6~Jy/beam.  Therefore, a constraint can be placed upon the line-of-sight column
density of the
OH in supernova remnants.  Assuming a spatially uniform background continuum filling the beam plus
a compact foreground OH masing column, filling a fraction of the
beam, the observed
intensity of a particular line can be written as

\begin{equation}
I_\nu =
(1-f)I_{0\nu}+f\{I_{0\nu}\,\mathrm{e}^{ -\tau_\nu }+B_\nu(T_{ex})[1-\mathrm{e}^{ -\tau_\nu }]\}\,,
\label{eqn:intensity}
\end{equation}  

where $I_{0\nu}$ is the background continuum intensity, $B_\nu$ is the
Planck function, $\tau_\nu$ is the optical depth, $f$ is the filling factor, and $T_{ex}$ is the
excitation temperature for the line.  Assuming that
$|T_{ex}| \gg 0.08$~K and the Rayleigh-Jeans approximation holds, equation (\ref{eqn:intensity}) can be rewritten as

\begin{equation}
T_b-T_{0}=f(T_{ex}-T_{0})(1-\mathrm{e}^{-\tau_\nu}).
\label{eqn:Temp}
\end{equation}

\noindent $T_b-T_0$ is the brightness temperature of the maser line detected by
the telescope.  Fixing the H$_2$ density at 10$^5$~cm$^{-3}$, a typical value
determined for 1720 MHz OH maser-emitting regions
\citep{1999ApJ...511..235L}, and using the model described in section
\ref{sec:motivation}, $\tau_\nu$, and $T_{ex}$ can be computed for a
particular $N_{OH}$.  For $N_{OH} \lesssim 10^{17.3}$~$\persqcm$, $|T_{ex}| \lesssim 1$~K,
which is much less than the continuum brightness temperatures typically found
in supernova remnants ($\sim$20~K, e.g. G349.7+0.2, \citealt{2004astro.ph..9302L}).  

Therefore, assuming that $T_{ex}$ is small, equation (\ref{eqn:Temp}) can be approximated by

\begin{equation}
\mathrm{e}^{-\tau_{\nu}}\approx 1+\frac{T_b-T_0}{fT_0}.
\end{equation}

As no 6049~MHz masers were detected, $T_b-T_0$ is the observed
3-$\sigma$ brightness temperature limit for the fields, which varies
from $\sim 0.3$ to 0.6~K ($1\, \mathrm{Jy/beam}= 0.803$~K). Therefore it can be seen that
the limits on the optical depth depend on the values for the filling factor and the
background continuum.  

\citet*{2008arXiv0802.3878H} found that a typical filling factor for
the 1720~MHz OH masers observed with the Green Bank Telescope (GBT) was
0.005.  As our beamsize is 3.4~arcmin compared to the 7.2~arcmin of
the GBT, the typical filling factor is multiplied by four to give an
appropriate value of 0.02.  Assuming the stated continuum brightness
temperature above of 20~K, it can be seen that 

\begin{equation}
\mathrm{e}^{-\tau_\nu}\sim 2\,,
\end{equation}  

\noindent and therefore $\tau_\nu \sim -1$.  From Fig. \ref{fig:tau},
this indicates a maximum OH column density of
$\sim 10^{16.5}$~$\persqcm$ at 6049~MHz. 

\begin{figure}
 \includegraphics[width=84mm]{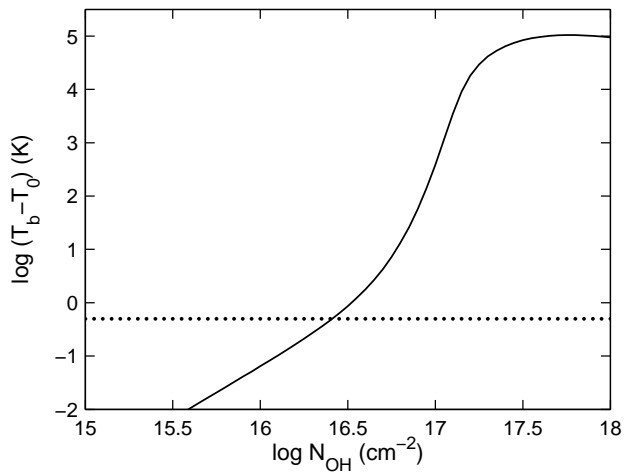}
                                
  \caption{Maser brightness temperature as a function of $N_{OH}$ for
    T=50~K for the 6049 MHz line of OH, assuming a filling factor of
    0.02 (see text).  The background continuum was
    assumed to have a brightness temperature of 20~K.  The horizontal
    dotted line marks the 3-$\sigma$ rms limit of our observations.}\label{fig:modellingresults}
\end{figure}

By way of example, consider the supernova remnant G349.7+0.2.  For this remnant, no 6049~MHz masers were detected to a 3-$\sigma$ level of
0.5~K.  Assuming the filling factor of 0.02 and background continuum
brightness temperature of 20~K from above, the predicted brightness
temperature for a 6049~MHz maser versus the column density is given in
Fig. \ref{fig:modellingresults}.   The minimum observed brightness
temperature of 0.5~K is also plotted, and the intersection of this line with
the predicted brightness temperature provides a limit on the column density
of the OH, consistent with the estimate from above.  It can be seen
that changing the filling factor and the
continuum brightness temperature does not make much difference to the
limit on $\tau_\nu$, and hence on the OH column density. 

If the column density in the supernova remnant was above 10$^{16.4}$~cm$^{-2}$,
6049 MHz maser emission should have been detected.  One possible
interpretation of this limit, therefore, is that the column density
must be below 10$^{16.4}$~cm$^{-2}$ for this remnant.  However, this
conflicts with results obtained by \citet{2008arXiv0802.3878H}
who reported a column density of 10$^{17}$~$\persqcm$ for areas in this
remnant.  

Assuming all other parameters are static, if the filling factor is an order of magnitude smaller than in the above calculations, the limit on the column
density is $\sim 10^{16.7}$~$\persqcm$ and if it is an order of magnitude
larger, the limit is $\sim 10^{15.9}$~$\persqcm$. The 3-$\sigma$ rms
brightness temperature varies from 0.27~K to 0.59~K over the fields observed, which gives column densities ranging
between $10^{16.3}$ and $10^{16.5}$~$\persqcm$.  Changes in this
parameter are therefore not significant for the fields
observed.  Changing the temperature $T$ of the gas does not appear to
significantly alter the results, as an increase to 75~K, as estimated by the modelling of
\citet{2008arXiv0802.3878H}, lowers the column density limit only
slightly to $\sim 10^{16.2}$.

We now consider the range of filling factors inferred from the 1720~MHz observations by
\citet{2008arXiv0802.3878H}.  The minimum filling
factor, $f=0.0005$, was obtained for clumps E, F in W28, with $\log
N_{OH}=17.14\pm 0.01$ for a kinetic temperature of 75~K.  The
corresponding factor for the Parkes beam is $f=0.002$, and for $T_0=20$~K, our column density limit was
$\sim10^{16.8}$.  Increasing $T_0$
reduces the limit, whereas, even at $T_0=2$~K, the limit on $N_{OH}$ is
still below the column density predicted by
\citet{2008arXiv0802.3878H}.  At the other extreme, the maximum filling factor, $f=0.137$,
is for clumps D,E,F in W44, giving a predicted $\log N_{OH}=15.95 \pm
0.03$ for a kinetic temperature of 50~K.  For our beam, $f=0.548$, and
for $T_0=20$~K, our column density limit was $\sim10^{15.7}$.  As
above, increasing $T_0$ reduces the limit.  If $T_0=10$~K, the column
density limit is still below the column density predicted by
\citet{2008arXiv0802.3878H}, but if it is reduced further to
$T_0=5$~K, the column density limit is above the predicted column
density value.  Therefore, for almost all cases, the
sensitivity of the observations was sufficient to detect 6049~MHz OH
maser emission that theoretically should have been produced.   

This indicates that, assuming that the column densities estimated by
\citet{2008arXiv0802.3878H} are correct, 6049~MHz OH maser emission should
have been observable in at least several of the remnants.  Several explanations may account for this conflict: the
filling factor may be much smaller than estimated because the background continuum
temperature is much higher, the continuum level could vary
significantly over
the beam, or there may be some other mechanism that is disrupting the
formation of the masers, such as a too high velocity gradient in the
OH column.  However, as 1720~MHz OH masers are produced in some of these
remnants and the excitation modelling does correctly predict the column
density ranges expected for the amplification of 1720~MHz OH masers,
any mechanism that stops the formation of the 6049~MHz OH masers
cannot affect the production of 1720~MHz OH masers.   Previous modelling by \citet{1999ApJ...511..235L} and \citet{2007IAUS..242..336W} only includes the thermal FWHM of the 1720~MHz and 6049~MHz OH lines,

\begin{equation}
\Delta v_T= \sqrt{\frac{8kT\log2}{17m_H}}
\end{equation}

\noindent where $k$ is Boltzmann's constant, $T$ is the kinetic temperature and $m_H$ is the mass of a hydrogen atom.  For $T=50$~K, this is 0.37~km~s$^{-1}$.  If we allow non-thermal broadening, as expected in clouds, for $\nH=10^5$~$\percc$ and $T=50$~K at an OH column density of $10^{17}$~$\persqcm$, the 6049~MHz inversion disappears at a FWHM linewidth of 0.54~km~s$^{-1}$, whereas 1720~MHz is no longer produced at 0.70~km~s$^{-1}$.  Therefore an increased sensitivity to velocity coherence may provide a partial explanation as to why 6049~MHz OH masers were not seen and yet 1720~MHz OH masers were seen in some remnants.  However, the observed line widths of 1720~MHz OH masers are generally wider than predicted by any of the current models.  

Note, however, that the OH column density expected in
SNR--cloud interactions is uncertain.  Existing models of OH
production rely on UV or X-ray dissociation of the water produced in
molecular shocks, and yield column densities $\la 10^{16}\persqcm$
\citep{1999ApJ...511..235L,1999ApJ...525L.101W}.  As the dissociation rate of OH
is approximately half that of H$_2$O, these models predict
OH/H$_2\mathrm{O}\la 2$, and typically much less.  This conflicts with
recent absorption measurements in IC 443, which find that
$N_\mathrm{H_2O}\sim 10^{14-15}\persqcm$ \citep{2005ApJ...620..758S} and
$N_\mathrm{OH}\sim 10^{16-17}\persqcm$ \citep{2006ApJ...652.1288H}.

\section{Conclusions}

OH excitation
calculations show that for column densities in excess of
10$^{17}$~$\persqcm$, maser emission at 6049~MHz may take over the role
of 1720~MHz masers in indicating a SNR~--~molecular cloud interaction.
The OH column density typically produced in these interactions is
uncertain, with densities of 10$^{17}$~$\persqcm$ calculated from
absorption measures in supernova remnants such as IC~443 \citep{2006ApJ...652.1288H}. 
A survey of 35 supernova remnants, a star-forming region and four fields in the Large and Small Magellanic
Clouds at 6049, 6030 and 6035~MHz was conducted using the Parkes
Methanol Multibeam receiver on the Parkes radio telescope. 

Eleven 6035 and three 6030~MHz OH masers associated with star-forming regions
were detected.  Two new 6035~MHz masers, one with associated 6030~MHz
maser emission, were found.  Where possible, magnetic fields were
derived from the Zeeman splitting of the polarisations.

No 6049~MHz maser emission was detected in any of the fields to a
3-$\sigma$ level
of 0.3\,--\,0.6~Jy/beam.  This places a limit on the column density of the OH
in the supernova remnants.  

For a typical supernova remnant such as G349.7+0.2, this limit is $10^{16.4}$~$\persqcm$,
indicating that if the column density of OH was above this value, 6049~MHz maser emission would have been observed.  However, this
conflicts with results obtained by \citet{2008arXiv0802.3878H} that
indicate that the column density is $10^{17}$~$\persqcm$ in sections
of the remnant where 1720~MHz OH masers are found.  Several explanations could account for this
discrepancy, such as the filling factor accounting for the size of the
maser spot compared to the beam being too high or the continuum
temperature being more variable over the beam than anticipated, but
are not likely.  Another possibility is that the velocity coherence is too low under some conditions for the 6049~MHz OH masers to be produced, while still high enough for 1720~MHz OH masers to be produced.

\section*{Acknowledgments}

We would like to thank Catherine Braiding for helping with the
observing, Stacy Mader for observing assistance and help with data reduction, and James Caswell for
assistance with maser identification.

\appendix

\bsp

\label{lastpage}

\end{document}